\documentclass[final,1p,times]{elsarticle} % do not change
\usepackage{graphicx} % do not change
\usepackage{amssymb} % do not change
\usepackage{amsthm} % do not change
\journal{Nuclear Physics A} % do not change
\begin{document} % do not change

\begin{frontmatter} % do not change

%% QM09Author: please enter your
%% Title, author and address info here; please do not use footnotes

% Your Title - please modify
\title{Jet energy loss and high $p_T$ photon production in hot quark-gluon plasma}

% Principle author, and co-authors - please modify
\author{G.-Y. Qin$^{a}$, C. Gale$^b$, S. Jeon$^b$, G. D. Moore$^b$, J. Ruppert$^b$}

% Address - please modify
% note that if you have authors from several institutions, we recommend
% labelling these [a], [b], [c] etc.
\address[a]{Department of Physics, The Ohio State University, Columbus, OH, 43210,
USA}
\address[b]{Department of Physics, McGill University, Montreal, Quebec, H3A 2T8, Canada}

\begin{abstract} % do not change
%% Text of abstract goes here - please modify

Jet-quenching and photon production at high transverse momentum
are studied at RHIC energies, together with the correlation
between jets and photons. The energy loss of hard partons
traversing the hot QGP is evaluated in the AMY formalism,
consistently taking into account both induced gluon emission and
elastic collisions. The production of high $p_T$ photons in Au+Au
collisions is calculated, incorporating a complete set of
photon-production channels. Putting all these ingredients together
with a (3+1)-dimensional ideal relativistic hydrodynamical
description of the thermal medium, we achieve a good description
of the current experimental data. Our results illustrate that the
interaction between hard jets and the soft medium is important for
a complete understanding of jet quenching, photon production, and
photon-hadron correlations in relativistic nuclear collisions.
\end{abstract} % do not change

\end{frontmatter} % do not change

%% QM09: we keep linenumbers at least for initial version
%\linenumbers % do not change

%% start of main text - please modify. Below is a sub-set (single section)
%% of an earlier proceedings that show how one can handle references
%% and figures etc.
%%\section{}\label{}

\section{Introduction}

Jet-quenching is one of the most important discoveries of the
Relativistic Heavy Ion Collider (RHIC) at Brookhaven National
Laboratory (BNL) \cite{Adcox:2001jp,Adler:2002xw}. Owing to the
strong interaction with the hot quark-gluon plasma (QGP) created
in these collisions, high transverse momentum ($p_T$) partons
produced from early hard scatterings suffer energy loss in the
medium, leading to significant suppression of high-$p_T$ hadrons
in central A+A collisions in comparison with those from 
binary-scaled p+p collisions \cite{Gyulassy:1993hr}. There has
been a lot of effort devoted to understanding the energy loss
experienced by hard jets in excited hadronic matter (e. g., see
Ref. \cite{Bass:2008rv}).

In addition to single-particle observables, more insight may be
gained through correlation studies, i.e., measuring high-$p_T$
hadron production associated with a high-$p_{T}$trigger. One
motivation is that correlation measurements put tighter
constraints on the initial momentum distribution of the partons
that fragment into the observed hadrons. In this context, high
$p_T$ photons have been considered as promising trigger particles
\cite{Wang:1996yh} as they are mostly produced from early hard
binary scatterings. Triggering on such photons should fix the
transverse momentum of the away-side parton  \cite{Renk:2006qg,
Arleo:2007qw, Zhang:2009rn}. However, it should be noted that
other photon sources, such as those involving jet-plasma
interactions, may contribute to photon-hadron correlations. We
present a study \cite{Qin:2009bk} of photon-hadron correlations at
RHIC which includes relevant high-$p_T$ photon-production
channels. The formalism developed in Ref. \cite{Qin:2007rn} is
employed to account consistently for collisional and radiative
energy loss of hard partons in the
Arnold-Moore-Yaffe (AMY) approach %\cite{AMY} %
\cite{Arnold:2001ms,Arnold:2001ba, Arnold:2002ja}. The thermalized
medium produced in Au+Au collisions is modeled by
(3+1)-dimensional hydrodynamics \cite{Nonaka:2006yn}.

\section{Calculation}

The evolution of the jet distributions $P(E,t) = {dN(E,t)}/{dE}$
in the medium is described by a set of coupled Fokker-Planck type
equations \cite{Jeon:2003gi,Turbide:2005fk}:
\begin{eqnarray}\label{FP-eq}
\frac{dP_j(E,t)}{dt} \!&=&\!  \sum_{ab}  \int  d\omega
\left[P_a(E+\omega,t) \frac{d\Gamma_{a\to
j}(E+\omega,\omega)}{d\omega dt}  - P_j(E,t)\frac{d\Gamma_{j\to
b}(E,\omega)}{d\omega dt}\right]. \ \ \ \ \ \
\end{eqnarray}
Here $d{\Gamma_{j\to a}(E,\omega)}/{d\omega dt}$ is the transition
rate, with $E$ the initial jet energy and $\omega$ the lost
energy. The radiative and collisional parts of the transition
rates have been discussed in Ref. \cite{Qin:2007zz,Qin:2007rn}.

To obtain high-$p_T$ hadrons produced in A+A collisions, the
medium-modified fragmentation function
$\tilde{D}_{h/j}(z,\vec{r}_\bot, \phi)$ is defined to take into
account the energy loss of jets in the medium:
\begin{eqnarray}
\label{mmff} \tilde{D}_{h/j}(z,\vec{r}_\bot, \phi) \!&=&\!\!
\sum_{j'} \!\int\! dp_{j'} \frac{z'}{z} D_{h/j'}(z')
P(p_{j'}|p_j,\vec{r}_\bot, \phi). \ \ \ \
\end{eqnarray}
Here $z = p_h / p_{j}$ and $z' = p_h / p_{j'}$, with $p_h$ the
hadron momentum and $p_{j}$($p_{j'}$) the initial (final) jet
momentum. $P(p_{j'}|p_j,\vec{r}_\bot, \phi)$ represents the
probability of obtaining a jet $j'$ with momentum $p_{j'}$ from a
given jet $j$ with momentum $p_j$ and is obtained by solving
Eq.~(\ref{FP-eq}). As the energy loss depends on the local medium
profiles along the jet path, one needs to convolve over the
distribution of jet production position $\vec{r}_\bot$ and
propagation direction $\phi$.

To calculate the spectrum of high-$p_T$ non-decay photons 
produced in relativistic nuclear collisions, one needs to take into account
all the important sources: early hard direct photons,
fragmentation photons, and jet-medium photons. Prompt direct photons are
mostly produced from early hard collisions between partons from
two initial nuclei, through quark-anti-quark annihilation and
quark-gluon Compton scattering. Fragmentation photons are produced
by the surviving high energy jets escaping the medium. As in
high-$p_T$ hadron production, one may define a medium-modified
photon fragmentation function
$\tilde{D}_{\gamma/j}(z,\vec{r}_\bot, \phi)$. Jet-medium photons
are produced during the passage of high energy jets through the
nuclear medium via induced photon bremsstrahlung and jet-photon
conversions. This defines an photon evolution equation solved with
the jet evolution,
\begin{eqnarray}
\label{photon_evolve} \frac{dP^{\rm JM}_\gamma(E,t)}{dt} \!&=&\!
\int  d\omega P_{q\bar{q}}(E{+}\omega,t) \frac{d\Gamma^{\rm
JM}_{q\to \gamma}(E{+}\omega,\omega)}{d\omega dt}. \ \ \ \ \ \
\end{eqnarray}
Here ${d\Gamma^{\rm JM}_{q\to \gamma}}/{d\omega dt} =
{d\Gamma^{\rm brem}_{q\to \gamma}}/{d\omega dt} + {d\Gamma^{\rm
conv}_{q\to \gamma}}/{d\omega dt}$. The transition rates
$d\Gamma^{\rm brem}_{q\to \gamma}/d\omega dt$ for photon
bremsstrahlung processes are discussed in Ref. %\cite{AMY}
\cite{Arnold:2001ms, Arnold:2001ba, Arnold:2002ja} , and
jet-photon conversion rates $d\Gamma^{\rm conv}_{q\to
\gamma}/d\omega dt$ may be inferred from the photon emission rates
for those processes.

For photon-hadron correlations, one defines a yield per-trigger,
representing the momentum distribution of away-side hadrons given
a trigger photon in the near side with momentum $p_T^\gamma$:
\begin{eqnarray}
P(p_T^h|p_T^\gamma) = {P(p_T^h,p_T^\gamma)}/{P(p_T^\gamma)}.
\end{eqnarray}
Note that  $P(p_T^\gamma)$ is the single-particle $p_T$
distribution and $P(p_T^\gamma, p_T^{h})$ is the $\gamma$-$h$ pair
$p_T$ distribution. Often, a photon-triggered fragmentation
function is defined,
\begin{eqnarray}
D_{AA}(z_T,p_T^\gamma) = p_T^\gamma P_{AA}(p_T^h|p_T^\gamma),
\end{eqnarray}
with $z_T = p_T^h/p_T^\gamma$. The effect of the nuclear medium on
photon-hadron correlations may be quantified by the nuclear
modification factor $I_{AA}$ defined as,
\begin{eqnarray}
I_{AA}(z_T,p_T^\gamma) =
{D_{AA}(z_T,p_T^\gamma)}/{D_{pp}(z_T,p_T^\gamma)}.
\end{eqnarray}

\section{Results}

\begin{figure}[htb]
\begin{center}
\includegraphics[width=0.44\linewidth]{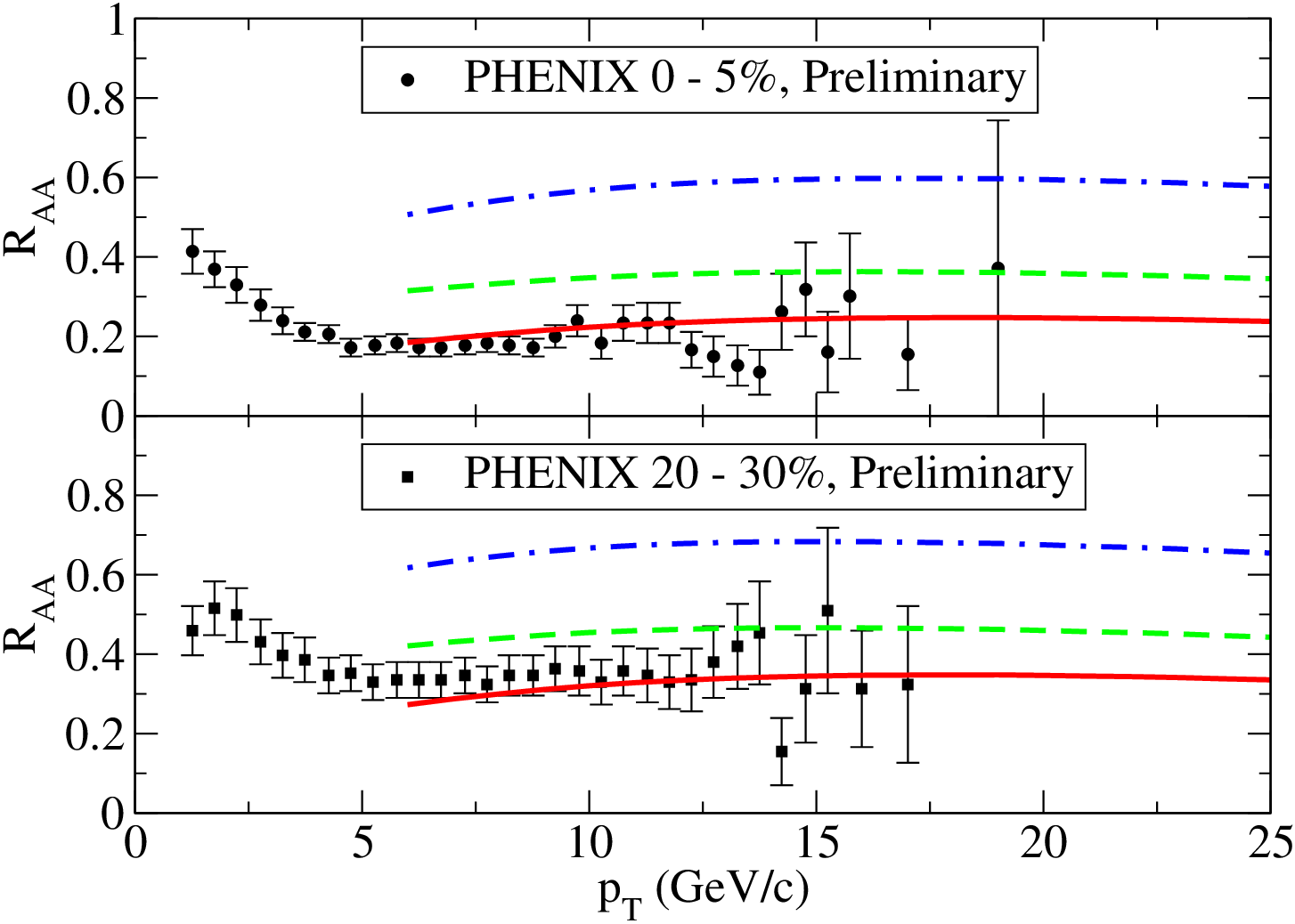}
\includegraphics[width=0.46\linewidth]{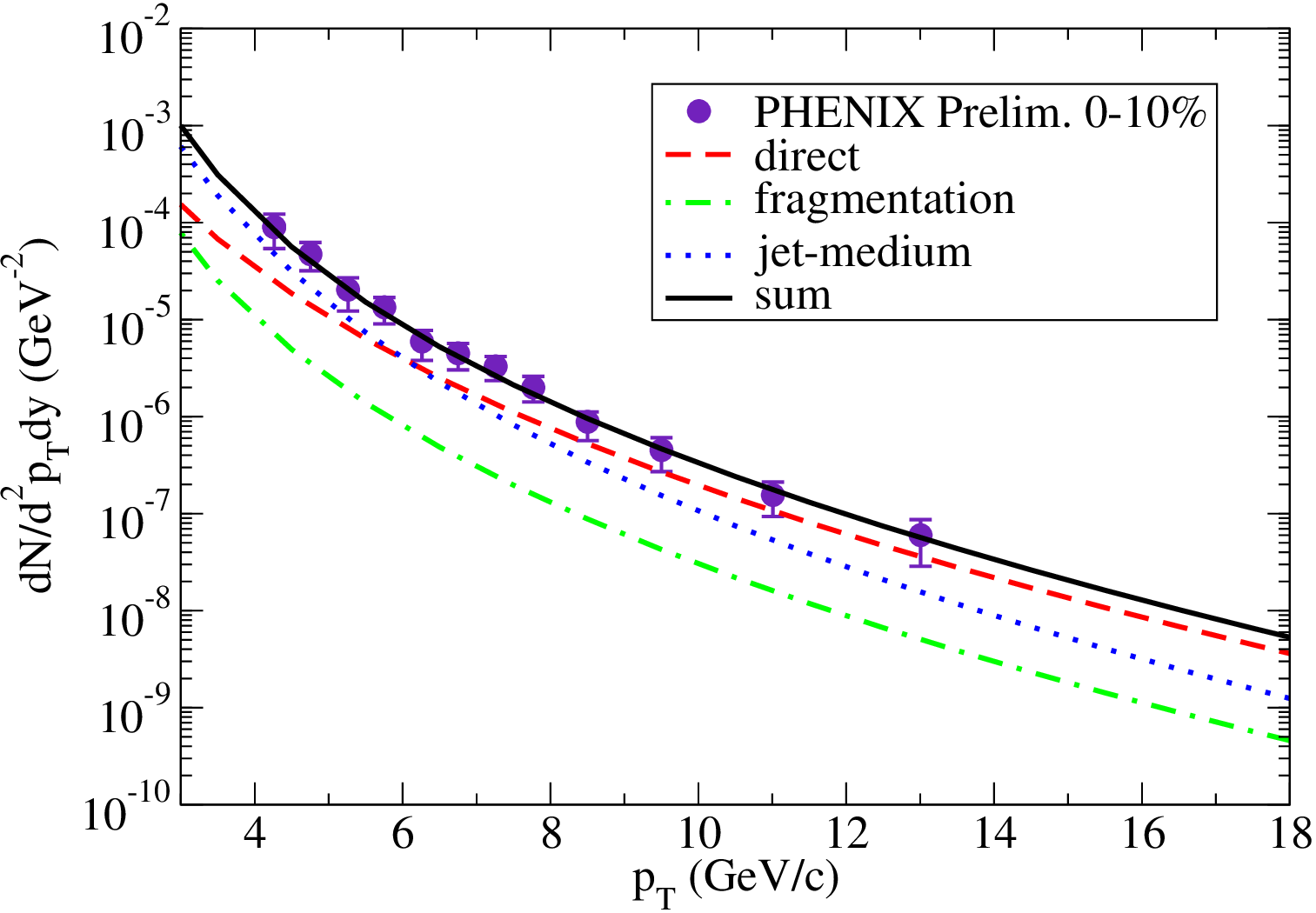}
\end{center} \caption{(Color online) Left: Neutral pion $R_{AA}$ in central and
mid-central Au+Au collisions at RHIC (from Ref. \cite{Qin:2007rn}, see text for explanation). Right: The
contributions from different channels to photon production in
most central Au+Au collisions at RHIC \cite{Qin:2009bk}. 
 } \label{raa}
\end{figure}

The results for jet-quenching and photon production in Au+Au
collisions at RHIC are shown in Fig. \ref{raa}. The left panel
shows neutral pion $R_{AA}$ measured at mid-rapidity for the most
central and mid-central collisions \cite{Qin:2007rn}. We also
compare the relative contributions of induced gluon radiations
(dashed) and elastic collisions (dash-dotted) to the final
$R_{AA}$ (solid). One finds that the overall magnitude of $R_{AA}$
is sensitive to both radiative and collisional energy loss. The
only free parameter, $\alpha_s$, -- the strong coupling constant --
is chosen such that the experimental measurement of $R_{AA}$ in
the most central collisions is described. The same value
$\alpha_s=0.27$ is then used throughout. In the right panel, we
show the relative contributions from different channels to
high-$p_T$ photon production in central collisions
\cite{Qin:2009bk}, and compare with PHENIX measurements
\cite{Isobe:2007ku}. While photons in the high-$p_T$ regime are
predominantly from early hard partonic collisions, the presence of
jet-medium interaction is nevertheless important to understand the
net photon production in Au+Au collisions at RHIC.

\begin{figure}[htb]
\begin{center}
\includegraphics[width=0.45\linewidth]{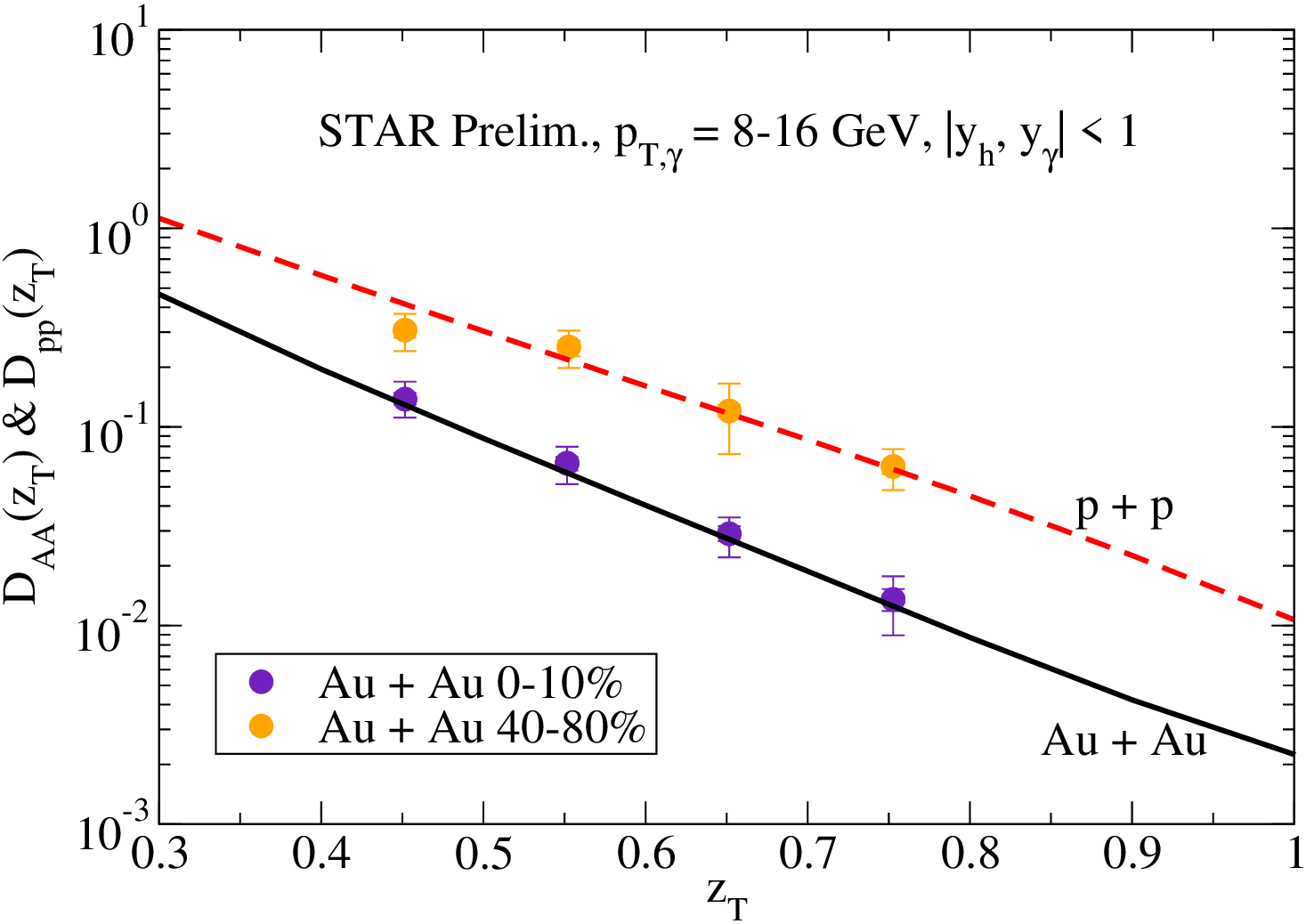}
\includegraphics[width=0.45\linewidth]{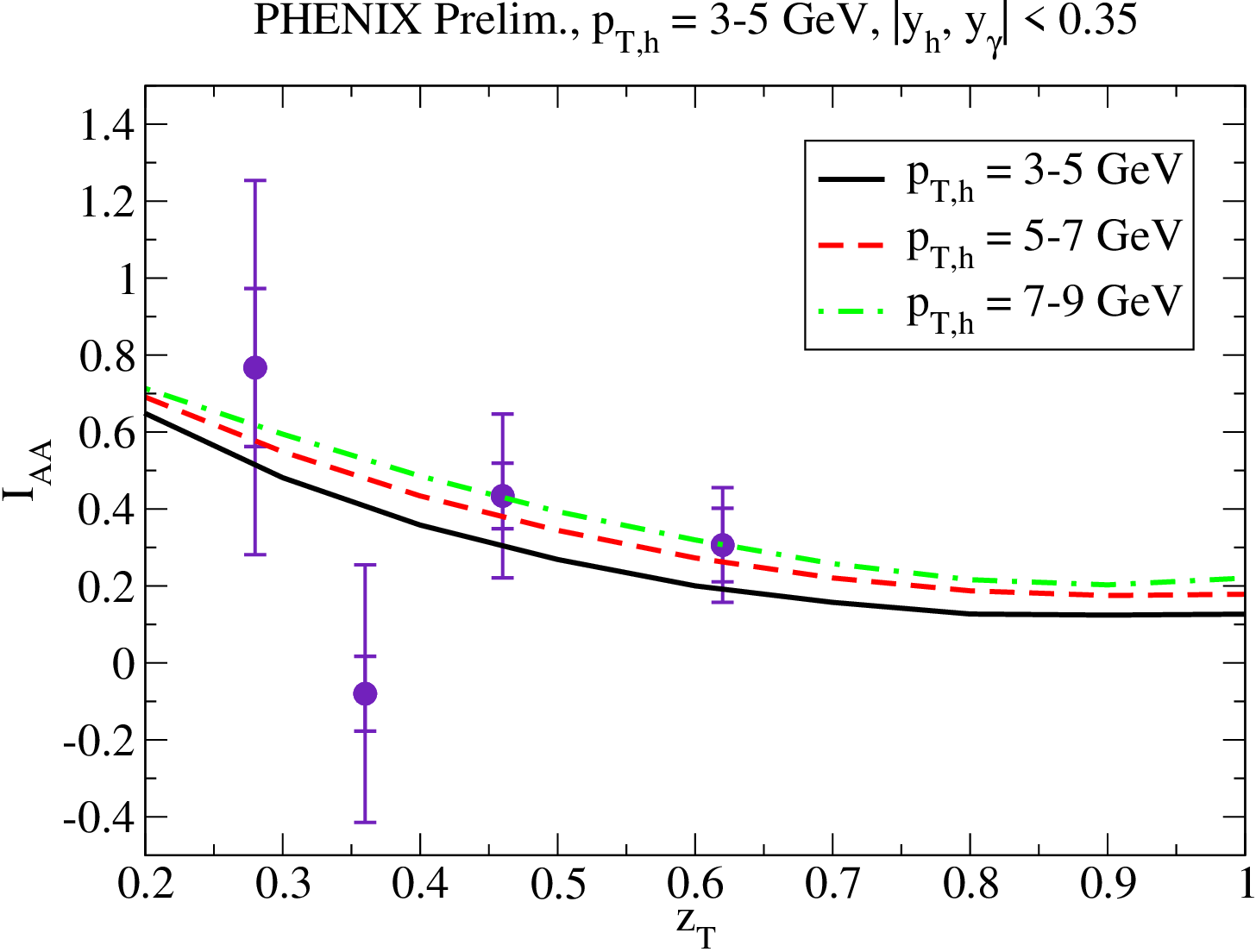}
\end{center}
\caption{(Color online) Photon-triggered fragmentation functions
and nuclear modification factor $I_{AA}$ as a function of $z_T$ in
Au+Au collisions compared to STAR (left) and PHENIX (right)
measurements. Figures are taken from Ref. \cite{Qin:2009bk}.}
\label{daa_iaa}
\end{figure}

In Fig. \ref{daa_iaa}, we show our results \cite{Qin:2009bk} for
photon-hadron correlations at high $p_T$ in Au+Au collisions at
RHIC. The left panel shows the photon-triggered fragmentation
function $D_{AA}(z_T)$ compared with STAR measurements
\cite{Hamed:2008yz}. The theoretical  photon-triggered
fragmentation function $D_{AA}(z_T)$ in central Au+Au collisions
agrees well with experimental data. Also the STAR measurements for
peripheral Au+Au collisions are consistent with p+p calculations.
In the right panel, we show the nuclear modification factor
$I_{AA}$ for photon-triggered hadron production compared with
PHENIX measurements \cite{Collaboration:2009vd}. While the fall of
$I_{AA}$ at low but increasing $z_T$ is due to the dominance of
direct photons, the flattening $I_{AA}$ predicted at higher values
of $z_T$ owes to the increasing influence of jet-medium photons
and of fragmentation photons. Also note that several bins of
hadron momenta are shown for comparison.

\section{Summary}

We have presented a study of jet energy loss, photon production
and photon-hadron correlations at high $p_T$ within a consistent
theoretical framework. Both induced gluon radiation and elastic
collisions are incorporated to calculate the energy loss of hard
jets traversing the hot and dense medium. A complete set of
photon-production channels is included in the computation of high
$p_T$ photon spectra. A fully (3+1)-dimensional hydrodynamical
evolution is employed to model the thermalized medium created at
RHIC. Our results have been shown to provide a good description of
the experimental measurements. This implies that jet-medium
interaction is important for jet-quenching, photon production, and
photon-hadron measurements at RHIC; the study of these
correlations will in turn provide insight on the detailed
structure of the excited medium created in high-energy nuclear
collisions.

%% end of main text

\section*{Acknowledgments} % please check/modify, comment out or delete if not needed

This work was supported in part by the U.S. Department of Energy
under grant DE-FG02-01ER41190, and in part by the Natural Sciences
and Engineering Research Council of Canada. G.-Y.Q. acknowledges
the support from the organizers of the Quark Matter 2009
conference.

%\bibliography{qin_reference_list}

\begin{thebibliography}{00} % do not change
\bibitem{Adcox:2001jp}
PHENIX, K.~Adcox {\em et~al.},
\newblock Phys. Rev. Lett. {\bf 88}, 022301 (2002), arXiv:nucl-ex/0109003.
%%CITATION = NUCL-EX/0109003;%%

\bibitem{Adler:2002xw}
STAR, C.~Adler {\em et~al.},
\newblock Phys. Rev. Lett. {\bf 89}, 202301 (2002), arXiv:nucl-ex/0206011.
%%CITATION = NUCL-EX/0206011;%%

\bibitem{Gyulassy:1993hr}
M.~Gyulassy and X.~nian Wang,
\newblock Nucl. Phys. {\bf B420}, 583 (1994), arXiv:nucl-th/9306003.
%%CITATION = NUCL-TH/9306003;%%

\bibitem{Bass:2008rv}
S.~A. Bass {\em et~al.},
\newblock Phys. Rev. {\bf C79}, 024901 (2009), arXiv:0808.0908.
%%CITATION = 0808.0908;%%

\bibitem{Wang:1996yh}
X.-N. Wang, Z.~Huang, and I.~Sarcevic,
\newblock Phys. Rev. Lett. {\bf 77}, 231 (1996), arXiv:hep-ph/9605213.
%%CITATION = HEP-PH/9605213;%%

\bibitem{Renk:2006qg}
T.~Renk,
\newblock Phys. Rev. {\bf C74}, 034906 (2006), arXiv:hep-ph/0607166.
%%CITATION = HEP-PH/0607166;%%

\bibitem{Arleo:2007qw}
F.~Arleo,
\newblock J. Phys. {\bf G34}, S1037 (2007), arXiv:hep-ph/0701207.
%%CITATION = HEP-PH/0701207;%%

\bibitem{Zhang:2009rn}
H.~Zhang, J.~F. Owens, E.~Wang, and X.-N. Wang,
\newblock (2009), arXiv:0902.4000.
%%CITATION = 0902.4000;%%

\bibitem{Qin:2009bk}
G.-Y. Qin, J.~Ruppert, C.~Gale, S.~Jeon, and G.~D. Moore,
\newblock (2009), arXiv:0906.3280.
%%CITATION = 0906.3280;%%

\bibitem{Qin:2007rn}
G.-Y. Qin {\em et~al.},
\newblock Phys. Rev. Lett. {\bf 100}, 072301 (2008), arXiv:0710.0605.
%%CITATION = 0710.0605;%%

\bibitem{Arnold:2001ms}
P.~Arnold, G.~D. Moore, and L.~G. Yaffe,
\newblock JHEP {\bf 12}, 009 (2001), arXiv:hep-ph/0111107.
%%CITATION = HEP-PH/0111107;%%

\bibitem{Arnold:2001ba}
P.~Arnold, G.~D. Moore, and L.~G. Yaffe,
\newblock JHEP {\bf 11}, 057 (2001), arXiv:hep-ph/0109064.
%%CITATION = HEP-PH/0109064;%%

\bibitem{Arnold:2002ja}
P.~Arnold, G.~D. Moore, and L.~G. Yaffe,
\newblock JHEP {\bf 06}, 030 (2002), arXiv:hep-ph/0204343.
%%CITATION = HEP-PH/0204343;%%

\bibitem{Nonaka:2006yn}
C.~Nonaka and S.~A. Bass,
\newblock Phys. Rev. {\bf C75}, 014902 (2007), arXiv:nucl-th/0607018.
%%CITATION = NUCL-TH/0607018;%%

\bibitem{Jeon:2003gi}
S.~Jeon and G.~D. Moore,
\newblock Phys. Rev. {\bf C71}, 034901 (2005), arXiv:hep-ph/0309332.
%%CITATION = HEP-PH/0309332;%%

\bibitem{Turbide:2005fk}
S.~Turbide, C.~Gale, S.~Jeon, and G.~D. Moore,
\newblock Phys. Rev. {\bf C72}, 014906 (2005), arXiv:hep-ph/0502248.
%%CITATION = HEP-PH/0502248;%%

\bibitem{Qin:2007zz}
G.-Y. Qin {\em et~al.},
\newblock Phys. Rev. {\bf C76}, 064907 (2007), arXiv:0705.2575.
%%CITATION = 0705.2575;%%

\bibitem{Isobe:2007ku}
PHENIX, T.~Isobe,
\newblock J. Phys. {\bf G34}, S1015 (2007), arXiv:nucl-ex/0701040.
%%CITATION = NUCL-EX/0701040;%%

\bibitem{Hamed:2008yz}
STAR, A.~M. Hamed,
\newblock J. Phys. {\bf G35}, 104120 (2008), arXiv:0806.2190.
%%CITATION = 0806.2190;%%

\bibitem{Collaboration:2009vd}
PHENIX, A.~Adare {\em et~al.},
\newblock (2009), arXiv:0903.3399.
%%CITATION = 0903.3399;%%

\end{thebibliography}
%\bibliographystyle{h-physrev5}

 % do not change
\end{document}